\documentclass[12pt]{article}
\usepackage{amsfonts}
\usepackage{amssymb}

\def\hybrid{\topmargin -20pt    \oddsidemargin 0pt
        \headheight 0pt \headsep 0pt
        \textwidth 6.35in       
        \textheight 9.25in       
        \marginparwidth .875in
        \parskip 5pt plus 1pt   \jot = 1.5ex}

\hybrid

\def\baselinestretch{1.2}

\catcode`\@=11

\def\marginnote#1{}
\newcount\hour
\newcount\minute
\newtoks\amorpm
\hour=\time\divide\hour by60
\minute=\time{\multiply\hour by60 \global\advance\minute by-\hour}
\edef\standardtime{{\ifnum\hour<12 \global\amorpm={am}%
        \else\global\amorpm={pm}\advance\hour by-12 \fi
        \ifnum\hour=0 \hour=12 \fi
        \number\hour:\ifnum\minute<10 0\fi\number\minute\the\amorpm}}
\edef\militarytime{\number\hour:\ifnum\minute<10 0\fi\number\minute}

\def\draftlabel#1{{\@bsphack\if@filesw {\let\thepage\relax
   \xdef\@gtempa{\write\@auxout{\string
      \newlabel{#1}{{\@currentlabel}{\thepage}}}}}\@gtempa
   \if@nobreak \ifvmode\nobreak\fi\fi\fi\@esphack}
        \gdef\@eqnlabel{#1}}
\def\@eqnlabel{}
\def\@vacuum{}
\def\draftmarginnote#1{\marginpar{\raggedright\scriptsize\tt#1}}

\def\draft{\oddsidemargin -.5truein
        \def\@oddfoot{\sl preliminary draft \hfil
        \rm\thepage\hfil\sl\today\quad\militarytime}
        \let\@evenfoot\@oddfoot \overfullrule 3pt
        \let\label=\draftlabel
        \let\marginnote=\draftmarginnote
   \def\@eqnnum{(\theequation)\rlap{\kern\marginparsep\tt\@eqnlabel}%
\global\let\@eqnlabel\@vacuum}  }

\def\preprint{\twocolumn\sloppy\flushbottom\parindent 2em
        \leftmargini 2em\leftmarginv .5em\leftmarginvi .5em
        \oddsidemargin -.5in    \evensidemargin -.5in
        \columnsep .4in \footheight 0pt
        \textwidth 10.in        \topmargin  -.4in
        \headheight 12pt \topskip .4in
        \textheight 6.9in \footskip 0pt
        \def\@oddhead{\thepage\hfil\addtocounter{page}{1}\thepage}
        \let\@evenhead\@oddhead \def\@oddfoot{} \def\@evenfoot{} }

\def\numberbysection{\@addtoreset{equation}{section}
        \def\theequation{\thesection.\arabic{equation}}}

\def\underline#1{\relax\ifmmode\@@underline#1\else
        $\@@underline{\hbox{#1}}$\relax\fi}

\def\titlepage{\@restonecolfalse\if@twocolumn\@restonecoltrue\onecolumn
     \else \newpage \fi \thispagestyle{empty}\c@page\z@
        \def\thefootnote{\fnsymbol{footnote}} }

\def\endtitlepage{\if@restonecol\twocolumn \else \newpage \fi
        \def\thefootnote{\arabic{footnote}}
        \setcounter{footnote}{0}}  

\catcode`@=12
\relax

\def\figcap{\section*{Figure Captions\markboth
        {FIGURECAPTIONS}{FIGURECAPTIONS}}\list
        {Figure \arabic{enumi}:\hfill}{\settowidth\labelwidth{Figure
999:}
        \leftmargin\labelwidth
        \advance\leftmargin\labelsep\usecounter{enumi}}}
 \relax
\def\tablecap{\section*{Table Captions\markboth
        {TABLECAPTIONS}{TABLECAPTIONS}}\list
        {Table \arabic{enumi}:\hfill}{\settowidth\labelwidth{Table
999:}
        \leftmargin\labelwidth
        \advance\leftmargin\labelsep\usecounter{enumi}}}
 \relax
\def\reflist{\section*{References\markboth
        {REFLIST}{REFLIST}}\list
        {[\arabic{enumi}]\hfill}{\settowidth\labelwidth{[999]}
        \leftmargin\labelwidth
        \advance\leftmargin\labelsep\usecounter{enumi}}}
 \relax
\makeatletter
\newcounter{pubctr}
\def\publist{\@ifnextchar[{\@publist}{\@@publist}}
\def\@publist[#1]{\list
        {[\arabic{pubctr}]\hfill}{\settowidth\labelwidth{[999]}
        \leftmargin\labelwidth
        \advance\leftmargin\labelsep
        \@nmbrlisttrue\def\@listctr{pubctr}
        \setcounter{pubctr}{#1}\addtocounter{pubctr}{-1}}}
\def\@@publist{\list
        {[\arabic{pubctr}]\hfill}{\settowidth\labelwidth{[999]}
        \leftmargin\labelwidth
        \advance\leftmargin\labelsep
        \@nmbrlisttrue\def\@listctr{pubctr}}}
 \relax
\makeatother
\newskip\humongous \humongous=0pt plus 1000pt minus 1000pt

\newif\ifdtup

\relax

\def\be{\begin{equation}}
\def\ee{\end{equation}}
\def\ba{\begin{eqnarray}}
\def\ea{\end{eqnarray}}

\def\del{\partial}

\def\r{\rho}
\def\a{\alpha}

\def\b{\beta}

\def\g{\gamma}
\def\G{\Gamma}
\def\d{\delta}

\def\e{\epsilon}

\def\th{\theta}

\def\m{\mu}
\def\n{\nu}

\def\Om{\Omega}
\def\l{\lambda}

\def\s{\sigma}

\def\cN{{\cal N}}

\def\bs{\bigskip}
\def\no{\noindent}

\def\qq{\qquad}

\def\IR{\relax{\rm I\kern-.18em R}}
\def\II{\relax{\rm 1\kern-.35em1}}

\def \ha {{1\over 2}}

\def \ov {\over}

\def\const{{\rm const.}}

\def\IR{\relax{\rm I\kern-.18em R}}
\def\inv{^{\raise.15ex\hbox{${\scriptscriptstyle -}$}\kern-.05em 1}}

\begin{document}

\newcommand{\beq}{\begin{equation}}
\newcommand{\eeq}[1]{\label{#1}\end{equation}}
\newcommand{\ber}{\begin{eqnarray}}
\newcommand{\eer}[1]{\label{#1}\end{eqnarray}}
\newcommand{\eqn}[1]{(\ref{#1})}
\begin{titlepage}
\begin{center}

\hfill NEIP-02-004\\
\vskip -.1 cm
\hfill hep--th/0205099\\

\vskip .5in

{\Large \bf Branes with fluxes wrapped on spheres}
\vskip 0.4in

{\bf Rafael Hern\'andez$^1$}\phantom{x} and\phantom{x}
 {\bf Konstadinos Sfetsos}$^2$ 
\vskip 0.1in

${}^1\!$
Institut de Physique, Universit\'e de Neuch\^atel\\
Breguet 1, CH-2000 Neuch\^atel, Switzerland\\
{\footnotesize{\tt rafael.hernandez@unine.ch}}

\vskip .2in

${}^2\!$
Department of Engineering Sciences, University of Patras\\
26110 Patras, Greece\\
{\footnotesize{\tt sfetsos@mail.cern.ch, des.upatras.gr}}\\

\end{center}

\vskip .3in

\centerline{\bf Abstract}
\no
Following an
eight-dimensional gauged supergravity approach 
we construct the most general solution 
describing D6-branes wrapped on a K\"ahler four-cycle taken 
to be the product of two spheres of different radii. 
Our solution interpolates between a Calabi--Yau four-fold 
and the spaces $S^2 \times S^2\times S^2\times \IR^2$ or 
$S^2 \times S^2\times \IR^4$, depending on generic choices for the parameters.
Then we turn on a background four-form field strength, corresponding 
to D2-branes, and show explicitly how our solution is deformed. For
a particular choice of parameters it represents a 
flow from a Calabi--Yau four-fold times the
three-dimensional Minkowski space-time in the 
ultraviolet, to the space-time $AdS_4 \times Q^{1,1,1}$ in the infrared.
In general, the solution in the infrared has a singularity which 
within type-IIA supergravity corresponds to the near horizon 
geometry of the solution for the D2-D6 system. 
Finally, we uncover the relation with work done in the eighties
on Freund--Rubin type compactifications.

\noindent

\vskip .4in
\noindent

\end{titlepage}
\vfill
\eject

\def\baselinestretch{1.2}


\baselineskip 20pt

Branes wrapped on supersymmetric cycles provide a natural path to obtain gravity duals of field theories 
with low supersymmetry. These field theories are twisted since 
preserving some supersymmetry after wrapping 
the brane, requires the identification (expressed better, the relation) 
of the spin connection on the cycle 
and some external R-symmetry gauge fields 
\cite{BSV}. Therefore,
the dual supergravity solutions can be naturally constructed in an 
appropriate gauged supergravity, and are eventually 
lifted to ten or eleven dimensions. 
This approach was started in \cite{MN}, 
and has been further developed for a wide variety 
of branes wrapped on diverse supersymmetric cycles \cite{AGK}-\cite{Gursoy}.
  
The case of D6-branes is of special interest because they lift to pure geometry in eleven dimensions. This fact allows 
to argue how compactifications of M-theory on manifolds with reduced holonomy arise as the local eleven 
dimensional description of D6-branes wrapped on supersymmetric cycles in manifolds of lower dimension with a 
different holonomy group \cite{Gomis}.
This extends the work of \cite{AMV},
where D6-branes wrapping the three-cycle in the deformed 
conifold were shown to be described in eleven dimensions as 
a compactification on a seven manifold with $G_2$ holonomy. These lifts 
to eleven dimensions for D6-branes wrapping various cycles were explicitly 
constructed using eight dimensional gauged 
supergravity \cite{Salam} in \cite{EN,Hernandez,GM,HS}.
  
However these purely gravitational geometries are deformed when background 
fluxes are included. 
In \cite{GM} M-theory on a Calabi-Yau four-fold was shown to
arise as the eleven dimensional description of D6-branes wrapped on
K\"ahler four-cycles inside Calabi-Yau three-folds.
The deformation of this background by a four-form field 
strength along the unwrapped coordinates was recently considered in 
\cite{Gursoy}, where it was shown to induce a flow 
from $\mathbb{E}_{2,1} \times CY_4$ at ultraviolet to $AdS_4 \times Q^{1,1,1}$ 
in the infrared limit. 
  
The four-cycle in \cite{GM,Gursoy} was taken to be a product of 
two two-spheres of the same 
radius so that the metric was Einstein. 
In this paper we will eliminate the Einstein condition on the four-cycle 
allowing the spheres to have different radii 
and will also introduce a four-form flux.
When lifted to eleven dimensions, and in the absence 
of flux, our solution will represent M-theory on a Calabi--Yau four-fold. 
We will find a three parameter 
family of metrics in which 
the conical singularity of the four-fold is generically 
resolved
by being replaced by a {\it bolt} or {\it nut} singularity which is removable
\cite{Gibbons}.
Then we turn on a background four-form field strength, corresponding to 
D2-branes, which provides another mechanism for resolving the 
singularity. After determining the most general supersymmetry preserving 
solution we discuss its behavior for various choices for the parameters. 
A special choice of parameters leads to an eleven-dimensional 
solution that flows from a Calabi--Yau four-fold times
the three-dimensional Minkowski space-time in the 
ultraviolet, to the space-time 
$AdS_4 \times Q^{1,1,1}$ in the infrared, where $Q^{1,1,1}$ is the 
seven-manifold coset space $SU(2)^3/U(1)^2$ 
that is supersymmetric \cite{Dauria}. While this is similar to 
\cite{Gursoy}, in the general case the singularity persists and 
is the same as in the near horizon metric for the D2-D6 system. 
Finally, we end the paper by making a precise connection of our
work with Freund--Rubin type compactifications 
of eleven-dimensional supergravity to four dimensions.

Before constructing our solution we will briefly review some relevant 
facts about gauged supergravity in eight dimensions which was 
constructed by Salam and Sezgin \cite{Salam}
through Scherk--Schwarz compactification of eleven-dimensional supergravity 
\cite{11sugra} on 
an $SU(2)$ group manifold. The field content of the theory 
consists of the metric $g_{\mu \nu}$, a dilaton 
$\Phi$, five scalars given by a unimodular 
$3 \times 3$ matrix $L_{\alpha}^{i}$ in the coset 
$SL(3, \IR)/SO(3)$ and an $SU(2)$ 
gauge potential $A_{\mu}$, all in the gravity sector, 
and a three-form coming from reduction of the eleven 
dimensional three-form.\footnote{Reduction of the eleven-dimensional 
three-form also produces a scalar, 
three vector fields and three two-forms. 
However, we will set all these fields to zero.} 
In addition, on the fermion side we have 
the pseudo--Majorana spinors $\psi_{\mu}$ and $\chi_i$.
    
The Lagrangian density for the bosonic fields 
is given, in $\kappa=1$ units, by
\ba
{\cal L} & = & \frac {1}{4} R - \frac {1}{4} e^{2 \Phi} F_{\mu \nu}^{\a}
F^{\mu \nu
\; \b} g_{\a \b}-
\frac {1}{4} P_{\mu \; ij} P^{\mu \; ij} - \frac {1}{2} (\partial_{\mu}
\Phi)^2 \nonumber \\
& - &
\frac {g^2}{16} e^{-2 \Phi} ( T_{ij} T^{ij} - \frac {1}{2} T^2)
- \frac {1}{48} e^{2\Phi} G_{\m \n \r \s} G^{\m \n \r \s} \ ,
\label{laad}
\ea
 where $F_{\mu \nu}^{\a}$ is the Yang--Mills field strength.
Supersymmetry will be preserved by bosonic solutions to the equations 
of motion of eight dimensional 
supergravity if the supersymmetry variations for the 
gaugino and the gravitino vanish. These are, respectively, given by 
\ba
\delta \chi_i & = &
 \frac {1}{2} (P_{\mu \; ij} + \frac {2}{3} \delta_{ij} \partial_{\mu} 
\Phi) \hat{\Gamma}^{j} 
\Gamma^{\mu} \epsilon - \frac {1}{4} e^{\Phi} F_{\mu \nu \; i} 
\Gamma^{\mu \nu} \epsilon \nonumber \\ 
               & - & \frac {g}{8} e^{-\Phi} (T_{ij} - \frac {1}{2} \delta_{ij} T) 
\epsilon^{jkl} \hat{\Gamma}_{kl} \epsilon - \frac {1}{144} e^{\Phi} G_{\m \n \r \s} \hat{\G}_i 
\G^{\m \n \r \s} \e  = 0 \ ,  
\label{susy}
\ea
and 
\ba
\delta \psi_{\gamma} & = & 
{\cal D}_{\gamma} \epsilon + \frac {1}{24} e^{\Phi} F_{\mu \nu}^{i} 
\hat{\Gamma}_i ( \Gamma_{\gamma}^{\: \: \mu \nu} 
- 10 \delta_{\gamma}^{\, \mu} \Gamma^{\nu}) \epsilon \nonumber \\ 
                     & - & \frac {g}{288} e^{- \Phi} \epsilon_{ijk} 
\hat{\Gamma}^{ijk} \Gamma_{\gamma} T \epsilon - \frac {1}{96} e^{\Phi} G_{\m \n \r \s} 
( \G_{\l}^{\: \: \m \n \r \s} - 4 \d^{\m}_{\: \: \l} \G^{\n \r \s} ) \e = 0 \ ,
\label{susyg}
\ea
where the covariant derivative is
\begin{equation}
{\cal D}_{\mu} \epsilon = \partial_{\mu} \epsilon + \frac {1}{4} \omega_{\mu}^{ab} \Gamma_{ab} 
\epsilon + \frac {1}{4} Q_{\mu \; ij} \hat{\Gamma}^{ij} \epsilon \, ,
\end{equation}
while $P_{\mu \, ij}$ and $Q_{\mu \, ij}$ are, respectively, 
the symmetric and antisymmetric quantities entering 
the Cartan decomposition of the $SL(3, \IR)/SO(3)$ coset, defined through
\begin{equation}
P_{\mu \, ij} + Q_{\mu \, ij} \equiv L_i^{\alpha} 
( \partial_{\mu} \delta_{\alpha}^{\, \beta} 
- g \, \epsilon_{\alpha \beta \gamma} A^{\gamma}_{\mu}) L_{\beta \; j} \, ,
\label{pq}
\end{equation}
and $T_{ij}$ is the $T$-tensor defining the potential energy associated to the scalar fields,
\begin{equation}
T^{ij} \equiv L^{i}_{\alpha} L^{j}_{\beta} \delta^{\alpha \beta} \, ,
\end{equation}
with $T \equiv T_{ij} \delta^{ij}$, and 
\be 
L_{\a}^{i} L^{\a}_j = \delta^{i}_j, \: \: \: \: L_{\a}^{i} L_{\b}^{j} \d_{ij} = g_{\a \b}, 
\: \: \: \: L^{i}_{\a} L^{j}_{\b} g^{\a \b} = \d^{ij}.
\ee
As usual, curved directions are labeled by greek indices, while 
flat ones are labeled by latin, and $\mu, a = 0,1, \ldots, 7$ are space-time coordinates, 
while $\alpha, i = 8,9,10$ are in the group manifold. Note also that upper indices in the gauge field, 
$A_{\mu}^{\alpha}$, are curved, and that the field strength in eight dimensional curved space is 
defined as
\be
G_{\m \n \r \s} = e^{- 4 \Phi/3} e^{a}_{\: \m} e^{b}_{\: \n} e^{c}_{\: \r} e^{d}_{\: \s} F_{abcd} \ .
\label{gef}
\ee
   
The $32\times 32$ gamma matrices in eleven dimensions can be represented as
\begin{equation}
\Gamma^{a} = \gamma^{a} \times \II_2, \: \: \: \: \: \: \: \: \hat{\Gamma}^{i} = \gamma_9 \times \tau^{i} \, ,
\end{equation}
where the $\g_a$'s denote the $16\times 16$ gamma matrices in eight dimensions 
and as usual $\gamma_9 = i \gamma^0 \gamma^1 \ldots \gamma^7$, 
so that $\gamma_9^2 = \II$, while $\tau^{i}$ 
are Pauli matrices. It will prove useful to introduce 
$\Gamma_9 \equiv \frac {1}{6i} \epsilon_{ijk} \hat{\Gamma}^{ijk} 
= -i \hat{\G}_1 \hat{\G}_2 \hat{\G}_3 = 
\gamma_9 \times \II_2$.
 
Let us now present the system under study in this paper and construct a solution describing this configuration. 
We will consider a D2-D6 brane system, with the D6-branes 
wrapped on a K\"ahler four-cycle inside a Calabi--Yau three-fold, 
and the D2-branes along the unwrapped directions. Keeping some supersymmetry unbroken involves an identification 
of the spin connection of the supersymmetric cycle and the gauge connection of the structure group of the normal 
bundle. When we wrap the D6-branes on the four-cycle, the $SO(1,6) \times SO(3)_R$ symmetry group of the branes in flat 
space is broken to $SO(1,2) \times SO(4) \times U(1)_R$. The breaking of the R-symmetry takes place because 
there are two normal directions to the D6-branes that are inside the 
Calabi--Yau three-fold; the R-symmetry is therefore 
broken to the $U(1)_R$ corresponding to them. The twisting will amount 
to the identification of this $U(1)_R$ with a $U(1)$ subgroup in one of the $SU(2)$ factors in 
$SO(4) \simeq SU(2) \times SU(2)$. The remaining scalar after the twisting, 
together with the vector and two 
fermions preserved by the diagonal group of 
$U(1) \times U(1)_R$, determine the field content of $\cN=2$ 
three-dimensional Yang--Mills. In the absence of D2-branes the lift to 
eleven-dimensions corresponds to M-theory on a Calabi--Yau 
four-fold \cite{Gomis,GM}. 

We will choose the four-cycle to be a product of 
two two-spheres of different 
radii, $S^2 \times {\bar S}^2$. 
The deformation on the world-volume of the D6-branes will then be described by 
a metric of the form
\be
ds_8^2 = e^{2f} ds_{1,2}^2  + d\r^2 + 
\a^2  d\Om_2^2 + \b^2  d{\bar\Om}_2^2\ ,
\label{metric}
\ee
where $ds_{1,2}^2$ is the three-dimensional Minkowski metric,  
the line elements for the two-spheres are 
\be 
 d\Om^2_2  
= d\th^2 +\sin^2 \th d\phi^2\ ,\qq
 d{\bar\Om}^2_2 = 
d{\bar \th}^2 +\sin^2 {\bar \th} d{\bar \phi}^2  \ ,
\label{sphee}
\ee
and $f$, $\a$ and $\b$ depend only on the radial variable $\r$. 
The same will be true for all 
additional fields that we will turn on. 
The four-form flux implied by the D2-branes along
the unwrapped directions will be 
\be
G_{x_0 x_1 x_2 \rho} = Q\: \frac {e^{-2 \Phi+3f}}{\a^2 \b^2} \ ,
\label{44ff}
\ee
where in the above $x_0,x_1,x_2$ are curved directions, $Q$
is a dimensionfull constant 
and the specific functional dependence is uniquely fixed
by the equation of motion for the three-form potential.

Turning now to the Killing spinor equation we should observe that it is quite useful to
introduce the triplet of Maurer--Cartan 1-forms on $S^2$
\be
\s_1 = \sin\th d \phi\ ,\qq  \s_2 =  d\th\ , \qq \s_3=\cos\th d\phi \ .
\ee
We note that they obey the conditions $d\s_i=\ha \e_{ijk} \s_j\wedge \s_j$,
so that they resemble the triplet of Maurer--Cartan forms on $S^3$, but 
obviously only two of them are the independent ones. We also introduce 
a similar
triplet ${\bar \s}_i$ defined on the other sphere ${\bar S}^2$. 

Consistency of the Killing spinor equations after splitting 
of the four-cycle into the product of spheres in (\ref{metric}) 
requires turning on only one of the components of the gauge field,
\be
A^3 =  - \frac {1}{g}(\s_3+{\bar \s}_3)\ , 
\label{wkjh}
\ee
thus realizing the breaking of the $SU(2)_R$ 
R-symmetry to $U(1)_R$ required by the twisting. 
In addition, consistency
requires turning on one of the scalars in $L^{i}_{\a}$,
\be
L^{i}_{\a} = \hbox {diag } ( e^{\l}, e^{\l}, e^{-2 \l}) \ ,
\ee
and imposing on the spinor the projections 
\ba
&&\G_7 \e =  - i \G_9 \e \ , \nonumber \\
&&\G_1 \G_2 \e =  \bar{\G}_1 \bar{\G}_2 \e = - \hat{\G}_1 \hat{\G}_2 \e \ .
\label{projections}
\ea
These leave in total four independent components for the spinor.
We note here that the simple relation \eqn{wkjh}, stating the equality (up 
to a constant) of the gauge field 
and the spin connection, is not valid in general when non-trivial 
scalar fields are present and instead it gets modified \cite{HS}.
  
With the above ansatz and projections on the spinor, the supersymmetry 
variations (\ref{susy}) and (\ref{susyg}) give the following conditions
\ba
\frac {d \Phi}{d \r} & = & \frac {g}{8} e^{- \Phi} (e^{-4 \l} + 2 e^{2 \l}) - \frac {1}{2g} e^{\Phi -2 \l} 
\left( \frac {1}{\a^2} + \frac {1}{\b^2} \right) - \frac {Q}{2} \frac {e^{-\Phi}}{\a^2 \b^2} \ , \nonumber \\
\frac {d \l}{d \rho} & = & \frac {g}{6} e^{-\Phi} ( e^{- 4 \l} - e^{2 \l} ) + \frac {1}{3g} e^{\Phi - 2 \l} 
\left( \frac {1}{\a^2} + \frac {1}{\b^2} \right) , \nonumber \\
\frac {1}{\a} \frac {d \a}{d \r} & = & \frac {g}{24} e^{- \Phi} ( 2 e^{2 \l} + e^{-4 \l} ) + 
\frac {1}{6g} e^{\Phi-2\l} \left( \frac {5}{\a^2} - \frac {1}{\b^2} \right) - 
\frac {Q}{2} \frac {e^{-\Phi}}{\a^2 \b^2} \ , 
\label{Killing}\\
\frac {1}{\b} \frac {d \b}{d \r} & = & \frac {g}{24} e^{- \Phi} ( 2 e^{2 \l} + e^{-4 \l} ) + 
\frac {1}{6g} e^{\Phi-2\l} \left( \frac {5}{\b^2} - \frac {1}{\a^2} \right) - 
\frac {Q}{2} \frac {e^{-\Phi}}{\a^2 \b^2} \ , \nonumber \\
\frac {d f}{d \r} & = & \frac {1}{3} \frac {d \Phi}{d \r} + \frac {2 Q}{3} \frac {e^{-\Phi}}{\a^2 \b^2} \ .
\nonumber
\ea
In addition we obtain a differential equation for the $\r$-dependence of the 
Killing spinor which can be easily integrated to yield
\be
\e=e^{f/2} \e_0\ ,
\ee
where $\e_0$ is a constant spinor subject to the projections \eqn{projections}.
In fact, this functional form of the Killing spinor can be deduced just from 
the supersymmetry algebra.

In order to solve the system \eqn{Killing}
we found it useful to redefine our variables as 
\ba
dr & = & e^{-\Phi/3} d \r \ , \qq A=f-\Phi/3\ ,
\nonumber \\
a_1  & = & \a \, e^{-\Phi/3} \ , \qq a_2 = \beta e^{- \Phi/3} \ , 
\label{reed} \\ 
a_3  & = & e^{\l + 2 \Phi/3} \ , \qq a = e^{- 2 \l + 2 \Phi/3} \ .
\nonumber
\ea
We also set for the rest of the paper the parameter $g=2$ since in any case 
in can be put to any value by appropriate rescalings.
Using the results of \cite{Salam}, the eleven-dimensional 
metric takes the form 
\be
ds_{11}^2 =  e^{2A} ds_{1,2}^2 +
 dr^2  + a_1^2 d\Om^2_2 +  a_2^2 d{\bar \Om}^2_2 + a_3^2 d{\hat \Om}^2_2  
+ a^2 \left( \hat{\s}_3 - \s_3 - \bar{\s}_3 \right)^2 \ ,
\label{lift}
\ee
where $\hat{\s}_i$ are left-invariant 
Maurer--Cartan $SU(2)$ one-forms 
satisfying as a triplet the conditions
$d \hat{\s}_i = \frac {1}{2} \e_{ijk} \hat{\s}_j \wedge \hat{\s}_k$.
We use the explicit representation
\ba
{\hat \s}_1 & = &
 \cos\hat \psi \sin\hat\th d \hat\phi\ -\sin\hat\psi d\hat \th
 \nonumber\\
 {\hat\s}_2 &  = &  \sin\hat \psi \sin\hat\th d \hat\phi\ 
+\cos\hat\psi d\hat \th\ ,
\label{ssss}\\ 
{\hat \s}_3 & = & d\hat \psi + \cos\hat\th d\hat\phi \ ,
\nonumber
\ea
so that the line element for the two-sphere $d{\hat \Om}^2_2$ is given by an 
expression similar to the ones in \eqn{sphee}. Using \eqn{gef} we may compute
the non-vanishing components of the four-form gauge field strength 
in eleven dimensions $F_{0127}$, where now all indices are in the tangent 
space. We easily find that
\be
F_{0127} = {Q\ov a_1^2 a_2^2 a_3^2 a }\ .
\label{12f}
\ee
Besides the metric and four-form, we may also use the fact that a Killing 
spinor can also be lifted from eight to eleven dimensions as
$\e_{11} = e^{-\Phi/6}\e=e^{A/2} \e_0$. 
Splitting the 32-component spinor $\e_{11}$
as $\e_{11}=\e_{1,2}\times \e_8$,
one can show that the projections \eqn{projections} leave 
two independent components in the 16-component spinor $\e_8$ whereas 
no restriction on the 2-component spinor $\e_{1,2}$ is needed. 
Hence we are left with $\cN=2$ supersymmetry in three dimensions.
We note that the proof of the above facts is completely parallel to the 
one we provided in our construction 
of $G_2$ holonomy manifolds from eight-dimensional 
gauged supergravity \cite{HS}, and this is the reason why we do not 
repeat it here.

After the redefinitions \eqn{reed},
the system \eqn{Killing} becomes 
\ba
a_1 \frac {da_1}{dr} & = & \frac {a}{2} - \frac {Q}{3} \frac {1}{a_2^2 a_3^2 
a} \ , \qq {\rm and\ cyclic\ in\ 1,2,3}\ ,
\nonumber \\ 
\frac {d a}{dr}  & = & 1- \frac {a^2}{2 } \left({1\ov a_1^2} 
+ {1\ov a_2^2} + {1\ov a_3^2}\right)
 - \frac {Q}{3} \frac {1}{a_1^2 a_2^2 a_3^2} \ ,
\label{ABCD}
\ea
whose solution determines the conformal factor as
\be
\frac {dA}{dr}  =   \frac {2 Q}{3} \frac {1}{a_1^2 a_2^2 a_3^2 a} \ .
\label{coonn}
\ee
  
\subsubsection*{The four-fold}
  
Let us first concentrate to the case where the D2-branes are absent, i.e.
when $Q=0$. Then the general 
solution to the system \eqn{ABCD} is 
\ba
&& a_1^2 =  R^2 + l_1^2, \qq a_2^2 = R^2 + l_2^2 \ , \nonumber \\
&& a_3^2 =  R^2, \qq a^2 = R^2 U^2(R) \ ,
\label{zero}
\ea
where
\be
U^2(R) = \frac {3 R^4 + 4 (l_1^2 + l_2^2) R^2 + 6 l_1^2 l_2^2+ 3 C/R^4}
{6 (R^2 + l_1^2 ) ( R^2 + l_2^2 )} \ .
\label{urr}
\ee
We also observe the relation of the two variables $r$ and $R$ via the 
differentials
\be
dr={2\ov U(R)} dR\ .
\label{drdr}
\ee
Here we have denoted three of the 
constants of integration by $l_1,l_2 $ and $C$
and we have absorbed the fourth one by an appropriate shift in the variable 
$R$.\footnote{Note that in the metric \eqn{lift} and in the system 
\eqn{ABCD} the three two-spheres are completely equivalent with no 
distinction between a ``space-time'' and an internal 
two-sphere. This equivalence seems to be broken by the solution \eqn{zero}.
However, this is only an artifact of setting the fourth 
integration constant to zero. The symmetry between the three 
two-spheres can be manifestly restored in the solution \eqn{zero} if
we make the variable shift $R^2\to R^2+l_3^2$ and simultaneously 
redefine $l_1^2\to l_1^2-l_3^2$ and $l_2^2\to l_2^2-l_3^2$.}
We can also see that in this case $e^{2\Phi}=R^3 U(R)$, $f=\Phi/3$ and
$A=0$.
Hence the lifted eleven dimensional solution in \eqn{lift} 
factorizes into the three-dimensional flat space-time 
and a Calabi--Yau four-fold, recovering the lift in \cite{Gomis} 
from $SU(3)$ holonomy in type-IIA string theory to 
$SU(4)$ holonomy in M-theory. Topologically the Calabi--Yau four-fold is a 
$\mathbb{C}^2$ bundle over $S^2 \times S^2$ and 
was also constructed with a different method in \cite{Cvetic} (for $C=0$).
This result includes those obtained in \cite{GM,Gursoy}, 
where both spheres in the four-cycle were taken to have the same radius, 
so that $l_1=l_2$. 

Let us note that for $R\to \infty$ the 
eight-dimensional metric takes the universal form
\ba
ds^2_8 \simeq   R^2 ( d\Om_2^2 + d{\bar \Om}^2_2
+ d{\hat \Om}^2_2)  
  + 8 dR^2 + \frac {R^2}{2} (\hat \s_3-\s_3-\bar \s_3)^2\ ,
\qq {\rm as}\quad R\to \infty\ .
\label{assy}
\ea
This solution is in fact exact for all values of $R$ since it can be obtained
from equations \eqn{zero}-\eqn{urr} by simply letting $l_1=l_2=C=0$.
However, extending it 
to the interior is problematic since we reach a singularity at $R=0$, 
where the fiber, the $S^2$ and the four-cycle collapse to a point. 
Resolving the singularity to avoid this collapse requires that we 
turn on some of the different
moduli parameters which also determine the behavior of the solution in the
interior.
In the following we further analyze the solution for different generic 
ranges of the
parameters and variables in order to determine for which ones the
singularity can indeed be resolved. 

We may systematically discuss the different cases as follows:

\no
\underline{$l_1=l_2=0$}: In this case, when the constant $C\geq 0$,
the variable $R\geq 0$ and then the 
manifold is singular at $R=0$. If, however, $C=-a^8<0$, where $a$ is a 
real positive constant, then the variable $R\geq a$.
Changing to a new radial variable $t=2\sqrt{a(R-a)}$ we find the 
behavior
\ba
ds^2_8 \simeq   a^2 ( d\Om_2^2 + d{\bar \Om}^2_2
+ d{\hat \Om}^2_2)  
  + \ dt^2 + t^2 (\hat \s_3-\s_3-\bar \s_3)^2\ ,\qq {\rm as}\quad t\to 0\ .
\label{bsoldd}
\ea
Therefore, near $t=0$ (or equivalently $R=a$) and for
constant $\th$ and $\phi$, as well as for the corresponding barred and 
hatted angles, the metric behaves as $dt^2 + t^2 d\hat \psi^2$
which shows that $t=0$ is a {\it bolt} singularity \cite{Gibbons}
which is removable provided that the periodicity 
of the angle $\hat{\psi}$ is restricted to $0\leq \hat \psi <2 \pi$.
Then the space becomes topologically 
$S^2\times S^2 \times S^2\times \IR^2$ and the full
solution interpolates between this space for $R\to a$ and the four-fold
\eqn{assy} for $R\to \infty$.

\no
\underline{$l_1^2> 0\ {\rm and}\ l_2^2> 0$}: 
In this case, when the constant $C>0$, the 
variable $R\geq 0$ and there is a singularity at $R=0$.
If, however, $C=0$ then we have the behavior
\ba
ds^2_8 \simeq   l_1^2  d\Om_2^2 + l_2^2 d{\bar \Om}^2_2 +
4 dR^2 + R^2 d{\hat \Om}^2_2 +
R^2 (\hat \s_3-\s_3-\bar \s_3)^2\ ,\qq {\rm as}\quad R\to 0\ .
\label{bsolt}
\ea
Hence, for constant $\th,\phi$ and $\bar\th,\bar\phi$ the metric 
behaves as $4 dR^2+ R^2(\hat\s_1^2 + \hat\s_2^2 + \hat\s_3^2) $ which shows
that we simply have a coordinate singularity in the polar coordinate system
on an $\IR^4$ centered at $R=0$. This is the so called {\it nut} singularity \cite{Gibbons}, 
which is removable by adding the point $R=0$ and changing to Cartesian 
coordinates. 
Therefore near $R=0$ the manifold becomes topologically 
$S^2\times S^2 \times \IR^4$. Then the full
solution interpolates between this space for $R\to 0$ and the four-fold
\eqn{assy} for $R\to \infty$.
If $C<0$ then there is an $R_0$ such that $U(R_0)^2=0$ 
(we take the largest root of this quartic, in $R_0^2$, equation) and therefore 
we have that $R\geq R_0$. 
Changing to a new radial variable $t=2\sqrt{R_0(R-R_0)}$ we find the 
behavior
\ba
ds^2_8 \simeq   (R_0^2+l_1^2) d\Om_2^2 + (R_0^2+l_2^2) d{\bar \Om}^2_2
+ R_0^2 d{\hat \Om}^2_2 
  + \ dt^2 + t^2 (\hat \s_3-\s_3-\bar \s_3)^2\ ,\: \: \: \: {\rm as}\: \: t\to 0\ .
\label{jkoldd}
\ea
Hence the behavior is similar to that found before in \eqn{bsoldd}, with a 
removable {\it bolt} singularity at $t=0$. Hence it will 
not be discussed any further.

\no
\underline{$l_1^2> 0\ {\rm and}\ l_2^2<0$ or 
$l_2^2< l_1^2 < 0$}: 
In this case it is convenient to define $\tilde l^2_2= -l_2^2$ so that 
$\tilde l_2^2>0$. Then, when 
$C< {1\ov 3} \tilde l_2^6(2 l_1^2+\tilde l_2^2)$, 
there is an $R_0>\tilde l_2$ obtained as
the largest root of the quartic (in $R_0^2$) equation $U^2(R_0)=0$ 
and therefore the variable $R\geq R_0$.
Then the behavior of the metric is given by \eqn{jkoldd} with the replacement
$l_2^2\to -\tilde l_2^2$ and as in that case, the {\it bolt}
singularity at $R=R_0$ is removable.
If $C= {1\ov 3} \tilde l_2^6(2 l_1^2+\tilde l_2^2)$, then $R_0\geq \tilde l_2$
and the metric near $R=R_0$ behaves as
\ba
ds^2_8 \simeq   (R_0^2-\tilde l_2^2) d\Om_2^2 + R_0^2 d{\hat \Om}^2_2 
+ dt^2 + {t^2\ov 4} \left(\bar\s_1^2 +\bar\s_2^2 + (\hat \s_3-\s_3-\bar \s_3)^2
\right)\ ,\qq {\rm as}\quad t\to 0\ ,
\label{jnutd}
\ea
where $t$ is some new radial variable. Hence, 
for constant $\th,\phi$ and $\hat\th,\hat\phi$ the metric 
behaves as $dt^2+{1\ov 4} t^2(\bar\s_1^2 + \bar\s_2^2 + \bar\s_3^2)$ which 
shows, as before in \eqn{bsolt}, that $t=0$ is a removable {\it nut}
singularity.
Finally, if $C> {1\ov 3} \tilde l_2^6(2 l_1^2+\tilde l_2^2)$, 
then $R\geq \tilde l_2$. We found that
in this case we have a curvature singularity at $R=\tilde l_2$.
We must note that in all of the above subcases the metric retains its
Euclidean
signature even for $l_1^2<0$ as long as $l_1^2+\tilde l_2^2>0$, which is in 
accordance with our original assumption.

\subsubsection*{Turning on the flux}
Dealing with fluxes is always a much more involved problem. 
However we will provide now a quite promising method that can 
probably be extended in general to settings similar to the one 
studied in this note. Turning to the system \eqn{ABCD} with $Q\neq 0$,
we first perform the transformations
\be
a_i= \tilde a_i e^{-Q x/3}\ ,\qq a= \tilde a e^{-Q x/3}\ ,
\ee
where the new functions $\tilde a_i$, $i=1,2,3$ and $\tilde a$ 
are to be determined
and $x$ is defined via the differential equation 
\be 
\frac {dr}{dx} = a_1^2 a_2^2 a_3^2 a =\tilde a_1^2 \tilde a_2^2 \tilde a_3^2 
\tilde a e^{-7 Q x /3}\ .
\label{drdx}
\ee
Then we can deduce from \eqn{ABCD} that the $\tilde a_i$'s and $\tilde a$
obey the system 
\ba
\tilde a_i \frac{d\tilde a_i}{d \tilde r} & = & \frac{\tilde a}{2} \ , 
\qq i= 1,2,3\ ,
\nonumber\\ 
\frac{d \tilde a}{d \tilde r}  & = & 
1- \frac{\tilde a^2}{2 } \left({1\ov \tilde a_1^2} 
+ {1\ov \tilde a_2^2} + {1\ov \tilde a_3^2}\right)\ ,
\label{ABCDt} 
\ea
where the new variable $\tilde r$ is related to $r$ 
via $d \tilde{r} = d r e^{Q x/3}$. 
This system is the same as the one in \eqn{ABCD}, but with $Q=0$. 
Hence we immediately conclude that the solution for the
$\tilde a_i$'s and $\tilde a$ is given by \eqn{zero}-\eqn{drdr} after the 
appropriate replacements of variables by the corresponding tilded ones.
In addition, we deduce from \eqn{coonn} that 
\be
A=f - \frac {\Phi}{3}   =  \frac {2}{3} Q x\ .
\ee
It remains to relate the variables $x$ and $R$, which is 
easily done using \eqn{drdx}. The result is better expressed via the integral 
\be
e^{-2 Q x}= -4 Q \int {dY \ov 
Y^2 \Big( Y^2 
+ \frac {4}{3} (l_1^2 + l_2^2) Y + 2 l_1^2 l_2^2 \Big) + C }\ +\ \const \ , 
\ee
where $Y=R^2$. Evaluating this integral is elementary, but the general result 
is not very illuminating. Instead, we will consider some limiting
cases. For $R\to \infty$, we have $e^{-2 Q x}\simeq 4 Q/3 R^{-6} +\const$.
Hence, $x$ tends to a constant, which without loss of generality can be chosen 
to be zero. Therefore for $R\to \infty$ our solution 
becomes the Calabi--Yau four-fold times $\mathbb{E}_{2,1}$. Towards the infrared 
the details of the solution depend on the various integration parameters.
We will consider first the case of $l_1=l_2=C=0$. Then the solution we 
gave above for $R\to \infty $ is actually valid for all values of $R$. 
Then for $R\to 0$ (and assuming $Q>0$) we have that $x\to -\infty $ and
we obtain the eleven-dimensional direct product 
solution $AdS_4\times Q^{1,1,1}$.
The case that we have just described was considered in \cite{Gursoy}.
When $C=0$, but for a generic choice of parameters $l_1$ and $l_2$
such that the radial variable $R\geq 0$, we have that $e^{-2 Q x}
\simeq {2 Q\ov l_1^2 l^2_2 R^2}$, as $R\to 0$. Then for $R=0$ there is a 
curvature singularity which however has a natural interpretation in terms of the
D2-D6 system. In order to see that, 
consider the type IIB supergravity solution 
obtained from dimensional reducing \eqn{lift} along the directions 
corresponding to the Killing vector field $\del/\del \hat \psi$. Then, for 
$R\to 0$ and 
after some algebraic manipulations, the metric becomes 
that for the D2-D6 system in the near horizon limit, 
with radial variable $r \sim R^2$ and  harmonic
functions $H_{2} , H_6\sim {1\ov r}$. It turns out that the constants 
$l_1^2$ and 
$l_2^2$ are naturally related to the ratio of D-brane charges.

\subsubsection*{Relation to Freund--Rubin compactifications}

In order to study the stability properties of Freund--Rubin type
compactifications \cite{FR}
{\it finitely} away from the supersymmetric vacua, a number of four-dimensional
supergravity actions have been constructed in the past by dimensionally 
reducing eleven-dimensional supergravity on seven-dimensional manifolds
representing deformations of the well known supersymmetric vacua. In this way
one obtains a four-dimensional theory of gravity coupled to scalars which 
model the deformations and which also
self-interact via a potential. This program was initiated by Page \cite{Page}
and further developed by Yasuda \cite{Yasuda}.
We think that it is important to make a precise connection with these 
works. We will restrict ourselves to the example considered in this paper,
but other cases can be worked out in a similar fashion.
In order to further dimensionally reduce eight-dimensional 
gauged supergravity on 
$S^2\times S^2$, we start with the more general, than 
\eqn{metric} and \eqn{44ff}, ansatz  
\be
ds_8^2 = e^{-4 h} G^{(4)}_{\m\n} dx^\m dx^\n + 
e^{2 h+ 2 \varphi}  d\Om_2^2 + e^{2 h-2 \varphi}  
d{\bar\Om}_2^2\ ,\qq \m,\n=0,1,2,7\ 
\label{metrge}
\ee
and 
\be
G_{x_0 x_1 x_2 x_7} = Q \: e^{-2\Phi-12 h}\sqrt{\det G^{(4)}}\ ,
\label{44fge}
\ee
where the functions $h$, $\varphi$ as well as the scalars $\Phi$ and $\l$
depend only on the variables of the four-dimensional metric $G^{(4)}_{\m\n}$.
and, as before, the four-form indices are curved.
Then, dimensionally reducing \eqn{laad}, which is the relevant part of the 
eight-dimensional gauged supergravity action of \cite{Salam},
we obtain a four-dimensional action for gravity coupled to 
scalars of the form
\be
S\sim \int d^4x \sqrt{\det G^{(4)}} \left({R^{(4)}\ov 4} + T - V \right)\ .
\label{aaac}
\ee  
Specifically, the kinetic term for the scalars is
\be
T = -3 (\del_\m h)^2- (\del_\m \varphi)^2  
-\ha (\del_\m \Phi)^2 -{3\ov 2} (\del_\m \l)^2 \ ,
\label{kkk1}
\ee
whereas their potential reads
\be
V  =   - e^{-6 h} \cosh2\varphi + {1\ov 4} e^{2 \Phi-4 \l-8 h} \cosh 4\varphi
+{1\ov 8} e^{-2\Phi-4 h} (e^{-8 \l} - 4 e^{-2 \l}) 
+{Q^2\ov 2} e^{-2 \Phi-12h}\ .
\label{poop}
\ee
This four-dimensional theory should be the same as that constructed 
in \cite{Yasuda} in relation to 
stability issues we mentioned 
of various compactifications of eleven-dimensional 
supergravity. In particular, eq.(19) of \cite{Yasuda} 
(for $p\!=\!q\!=\!r\!=\!1$) 
describes the 
four-dimensional theory obtained by a Freund--Rubin type compactification of 
eleven-dimensional supergravity on a deformed $Q^{1,1,1}$ space. 
In our normalization this action is of the form \eqn{aaac} with kinetic
term 
\be
T= -{63\ov 8} (\del_\m s)^2 -{21\ov 8} (\del_\m u)^2 -{3\ov 4} (\del_\m v)^2
-{1\ov 4} (\del_\m w)^2 \ 
\label{kkk2}
\ee
and potential\footnote{We differ from \cite{Yasuda} by a factor of 2 which
in not due to different normalization choices.}
\be
V= e^{9s}\left(- e^{u+v} \cosh w +{1\ov 4} e^{ 8 u + 2 v}\cosh 2w 
+ {1\ov 8} e^{8 u- 4v} -{1\ov 2} e^{ u -2 v} \right)
+ {Q^2\ov 2} e^{21 s}\ .
\label{ppo2}
\ee
We expect
that the action \eqn{aaac} with kinetic and potential terms \eqn{kkk1} and 
\eqn{poop} will be identical to the action with corresponding terms given 
by \eqn{kkk2} and \eqn{ppo2}.
Indeed, we found that the appropriate field redefinitions are
\be
w=2 \varphi\ ,\qq v={2\ov 3} (\l+\Phi - h)\ ,\qq u ={2\ov 21}
(2\Phi - 7 \l -2 h)\ ,\qq s=-{2\ov 7}(\Phi/3 +2 h)\ .
\ee

Let us finally note that the potential \eqn{poop} is derivable from 
the prepotential 
\be
W =- e^{-2 h-\Phi} (e^{2\l}+\ha e^{-4 \l}) - e^{\Phi-2 \l- 4h}\cosh 2 \varphi
+ Q e^{-\Phi - 6 h}\ ,
\ee
using the appropriate for four space-time dimensions 
formula
\be
V= {1\ov 8} \left[\left(\del W\ov \del \Phi\right)^2 + {1\ov 3}
\left(\del W\ov \del \l\right)^2 
+ {1\ov 6} \left(\del W\ov \del h\right)^2 + \ha 
\left(\del W\ov \del \varphi\right)^2 \right] 
- {3\ov 8} W^2\ .
\label{popp}
\ee
The system of equations \eqn{Killing} can also be obtained from the 
four-dimensional action \eqn{aaac} as follows. 
First we make a domain wall ansatz for the metric 
\be
ds_4^2 = e^{2 A_4} ds_{1,2}^2 + d\r_4^2 \ ,
\ee
where the conformal factor $A_4$ depends only on the fourth radial coordinate
$\r_4$ and the same is true 
for the scalars. Then, first order equations for the scalars can be 
obtained using 
\ba
&& {d \Phi\ov d\rho_4} = \ha {\del W\ov \del \Phi} \ ,\qq 
{d \l\ov d\rho_4} = {1\ov 6} {\del W\ov \del \l} \ ,
\nonumber\\
&& 
{d h\ov d\rho_4} = {1\ov 12} {\del W\ov \del h} \ ,\qq 
{d \varphi \ov d\rho_4} = {1\ov 4} {\del W\ov \del \varphi} \ ,
\label{ssii}
\ea
whereas the conformal factor is simply given by 
\be
{d A_4\ov d\r_4}=-\ha W\ .
\label{ssi1}
\ee
As usually, satisfying the above first order system implies that the 
second order equations of motion for the action \eqn{aaac}
are automatically satisfied as well.
We note at this point that we have constructed the prepotential $W$ in 
order to reproduce the expression for the potential $V$ using \eqn{popp}. 
This does not necessarily
guarantee that the first order equations \eqn{ssii} imply supersymmetry.
Nevertheless, this is true in our case. Indeed, 
taking into account the redefinitions 
$\a=e^{h+\varphi}$, $\b=e^{h-\varphi}$,
$f=A_4-2 h$ and $d\r=e^{-2h} d\r_4 $, we can verify that the system of 
\eqn{ssii} and \eqn{ssi1} coincides with the system \eqn{Killing}.

We note that the other cases that have been 
discussed in \cite{Gursoy} concerning the squashed $S^7$ and $N^{0,1,0}$
in the infrared, can be obtained from four-dimensional supergravity 
actions using the method that we outlined here. In addition, these actions 
coincide with those obtained in the past \cite{Page,Yasuda}. We also note that 
\cite{Yasuda} contains a few other cases worth a reinterpretation in terms of 
branes with fluxes wrapped on supersymmetric cycles.

The eleven dimensional description of solutions to eight
dimensional gauged supergravity corresponding to configurations of
wrapped branes in the presence of fluxes provides a fruitful tool
to approach compactifications of M-theory. In this note we have studied
in detail the case of D6-branes wrapped on K\"ahler four-cycles,
using as a representative example wrapping on $S^2\times \bar{S}^2$,
and proposed a clean method to construct solutions in the presence of
four-form field strength. We hope similar studies can be performed for
different configurations.


\bs\bs

\centerline {\bf Acknowledgments}

R. H. acknowledges the financial support provided through the European
Community's Human Potential Programme under contract HPRN-CT-2000-00131
``Quantum Structure of Space-time'' 
and by the Swiss Office for Education and Science 
and the Swiss National Science Foundation. 
K. S. acknowledges the financial support provided through the European
Community's Human Potential Programme under contracts 
HPRN-CT-2000-00122 ``Superstring Theory'' and HPRN-CT-2000-00131
``Quantum Structure of Space-time''.
He also acknowledges support by the Greek State
Scholarships Foundation under the contract IKYDA-2001/22, 
as well as NATO support
by a Collaborative Linkage Grant under the contract PST.CLG.978785.


\end{document}